\documentclass[aps,prl,twocolumn,groupedaddress,superscriptaddress]{revtex4-2}
\usepackage{graphicx}
\usepackage{amsmath}
\usepackage{epstopdf}
\usepackage{float}
\usepackage{xcolor}
\usepackage{times}
\usepackage{url}
\usepackage{subfigure}
\usepackage{siunitx} 
\usepackage{pgfplotstable} 
\usepackage{multirow}
\usepackage{booktabs}
\usepackage{longtable}
\begin{document}

	\title{Scaling law for three-body collisions near a narrow $s-$wave Feshbach resonance }
	
	
	\author{Jiaming Li}
	\email[]{lijiam29@mail.sysu.edu.cn}
	\affiliation{School of Physics and Astronomy, Sun Yat-Sen University, Zhuhai, Guangdong, China 519082}
	\affiliation{Center of Quantum Information Technology, Shenzhen Research Institute of Sun Yat-sen University, Shenzhen, Guangdong, China 518087}
	\affiliation{State Key Laboratory of Optoelectronic Materials and Technologies, Sun Yat-Sen University, Guangzhou, Guangdong, China 510275}
	
	\author{Shuai Peng}
	\affiliation{School of Physics and Astronomy, Sun Yat-Sen University, Zhuhai, Guangdong, China 519082}
	
	\author{Yirou Xu}
	\affiliation{School of Physics and Astronomy, Sun Yat-Sen University, Zhuhai, Guangdong, China 519082}
	
	\author{Shiyin Kuang}
	\affiliation{School of Physics and Astronomy, Sun Yat-Sen University, Zhuhai, Guangdong, China 519082}

	\author{Le Luo}
	\email[]{luole5@mail.sysu.edu.cn}
	\affiliation{School of Physics and Astronomy, Sun Yat-Sen University, Zhuhai, Guangdong, China 519082}
	\affiliation{Center of Quantum Information Technology, Shenzhen Research Institute of Sun Yat-sen University, Shenzhen, Guangdong, China 518087}
     \affiliation{State Key Laboratory of Optoelectronic Materials and Technologies, Sun Yat-Sen University, Guangzhou, Guangdong, China 510275}
     \affiliation{Quantum Science Center of Guangdong-Hongkong-Macao Greater Bay Area, Shenzhen, Guangdong, China 518048 }

	
\date{\today}
	
\begin{abstract}
Ultracold atomic gases provide a controllable system to study the inelastic processes for three-body systems, where the three-body recombination rate depends on the scattering length scaling. Such scalings have been confirmed in bosonic systems with various interaction strengths, but their existence with fermionic atoms remains elusive. In this work, we report on an experimental investigation of the scaling law for the three-body atomic loss rate $L_3$ in a two-component $^6$Li Fermi gas with the scattering length $a<0$. The scaling law is validated within a certain range of $a$ near the narrow $s$-wave Feshbach resonance, where $L_3\propto T|a|^{\num{2.60 +- 0.05}}$, and $T$ is the gas temperature. The scaling law is observed to have an upper and a lower bound in terms of the scattering length. For the upper bound, when $a\rightarrow \infty$, the power-law scaling is suppressed by the unitary behavior of the resonance caused by the strong three-body collisions. For the lower bound, $a\rightarrow 0$, the finite range effect modifies the scaling law by the effective scattering length $L_e$. These results indicate that the three-body recombination rate in a fermionic system could be characterized by the scaling law associated with the generalized Efimov physics.
\end{abstract}
	
	
	\maketitle
	
Studies of quantum three-body physics help to clarify some fundamental properties of a many-body system,  ranging from the formation of the first star~\cite{Cox1980} to the exotic nuclear dynamics~\cite{Muller1992nucl-th}. However, being immersed in a real system with many-particles, the three-body process is usually difficult to be solved in an analytical way. In past decades, ultracold quantum gases have provided an ideal experimental platform for quantitatively studying few-body physics, indebted to the precise-tuned interaction via Feshbach resonances~\cite{Chin2010RMP82.1225, Greiner2003Nature426.537,Jochim2003Science302.2101,Bourdel2004PRL93.050401,Karemer2006Nature440.7082,Scott2009science326.5960.1683,Wang2014NP10.768-773,Chen2022PRL128.153401}.
As a notable example, the dependence of the three-body loss rate $L_3$ on the two-body scattering length $a$ attracted substantial attention~\cite{Braaten2010PRA81.013605,Rem2013PRL110.163202,Wacker2016PRL117.163201,Johansen2017Nature13.731-735, Esry2001PRA65.010705(R)}. 

The $L_3\propto a^4$ scaling of three identical bosons have been theoretically predicted and experimentally confirmed in both $a>0$ and $a<0$ regions~\cite{KStamper-Kurn1998PRL80.2027, Weber2003PRL91.123201,Gross2009PRL103.163202}. For identical fermions, their $L_3$ relates to the $p-$wave scattering volume $V_p$ has been predicted to scale as $V_p^{8/3}$ when $V_p<0$~\cite{Greene2002PRL90.053202}, which was also confirmed in the experiment of a single-spin component $^6$Li fermi gas~\cite{Yoshida2018PRL120.133401}.
For a two-component Fermi gas with the $s-$wave interaction, a weakly bound molecule is formed in the $a>0$ regime where the predicted $L_3\propto a^6$ was verified recently using a homogeneous gas with a box trapping potential in the Bose-Einstein condensation (BEC) side of the broad Feshbach resonance of $^6$Li ~\cite{Petrov2003PRA67.010703R, Ji2022PRL129.203402}. 

On the other hand, the scaling law in the Bardeen-Cooper-Schrieffer (BCS) side, $a<0$, has yet to be well understood. So far, in the BCS side, $L_3$ has been observed scaling as $|a|^{\num {0.79 +- 0.14}}$ in a quasi-two-dimensional Fermi gas~\cite{Du2009PRL102.250402}, which is quite different from the theoretical prediction of $|a|^{2.455}$ basing on the adiabatic hyperspherical representation with an effective three-body repulsive potential~\cite{DIncao2005PRL94.213201}. Very recently, Ref.~\cite{Chen2023PRA107.033329} proposes that the $p-$wave interaction between the two spin-up fermions in two-component Fermi gases is responsible to the discrepancy. Meanwhile, it is noted that the scaling behavior could be strongly influenced by the unitary behavior~\cite{DIncao2009JPBAMOP42.044016, Chen2023PRA107.033329} as well as the finite-range effect~\cite{Yoshida2018PRL120.133401, Ho2012PRL108.250401}. These considerations raise the interesting question of whether the predicted scaling in Ref.~\cite{DIncao2005PRL94.213201} exists only in a limit range of $a$.

From the experimental viewpoint, testing the scaling law using the $^6$Li narrow Feshbach resonance has both advantages and disadvantages compared to the broad resonance. The first advantage is that $a$ can be tuned from minus infinity all the way to the zero in the BCS side of the narrow resonance, providing a wide tuning range to avoid the unitary behavior. It is predicted that the scaling law on $|a|$ is only valid when the system is in the threshold regime $|a|<a_c$~\cite{DIncao2018JPBAMOP51.043001}, where $a_c=2\hbar^2/(3 m |r_{eff}| k_B T)$. $m$ is the atomic mass, $r_{eff}$ is the effect range of the resonance, $\hbar$ is the Planck constant, and $k_B$ is the Boltzmann constant.
At a temperature of a few micro-Kelvin, $a_c$ is about one thousand Bohr radius $a_0$ for $^6$Li, while the background scattering length is about -1405 $a_0$ in the broad Feshbach resonance, so it is difficult to obtain the regime of $|a|<a_c$~\cite{unitaryBFR}.
Second, three-body recombination is the dominant mechanism for atom loss near the narrow Feshbach resonance~\cite{Li2018PRL120.193402}, providing a more obvious way to determine the three-body loss rate. For disadvantage, the scaling law in the narrow Feshbach resonance is affected by the finite-range effect. Previous theoretical predictions of the scaling have been limited to the regime $|a|>r_{vdw}$, where $r_{vdw}$ denotes the range of the van de Waals potential.
If $|a|> r_{vdw}$ is not well satisfied, the finite-range effect must be included in the scaling models~\cite{Petrov2004PRL93.143201, Braaten2006PR428.259, Wang2011PRA83.042710}.

In this letter, we report the precision measurement of the three-body atomic loss rate $L_3$ for an ultracold two-component $^6$Li Fermi gas near its narrow $s-$wave Feshbach resonance.
We first confirm that the threshold behavior of $L_3\propto T$ is valid in the narrow $s-$wave Feshbach resonance and covers an extremely large range as long as the system is away from the unitary regime $|a|>a_c$.
We then study $L_3$ in the intermediate interacting regime $r_{vdw}<|a|<a_c$, and find that $L_3\propto|a|^{2.60(5)}$. 
Moreover, due to the finite range effect, a correction of $|a|^{5.9(15)}/|L_e|^{\num {3.3 +- 1.0}}$ is obtained when $|a|$ closes to $a_c$. Here $L_e=(1/2 r_{eff} a^2)^{1/3}$ is an effective scattering length including both $a$ and $r_{eff}$~\cite{Blackley2014PRA89.042701, Shotan2014PRL113.053202}.
This correction becomes larger as $|a|$ increases.
For the data near the zero crossing  $|a|<r_{vdw}$, we find that the scaling behavior  $L_3\propto (|a|^{2.60(5)}+L^{2.58(4)}_e)$, such correction of $L^{2.58(4)}_e$ becomes larger with decreasing $|a|$.
In the zero-crossing limit, the scaling dependence only on $L_e$. With a correction factor, we find that the value of $L_e$ closes to $r_{vdw}$, which is consistent with the fact that the only length scale in the zero-crossing limit is $r_{vdw}$.

\begin{figure*}[htbp]
	\begin{center}
		\includegraphics[width=2\columnwidth, angle=0]{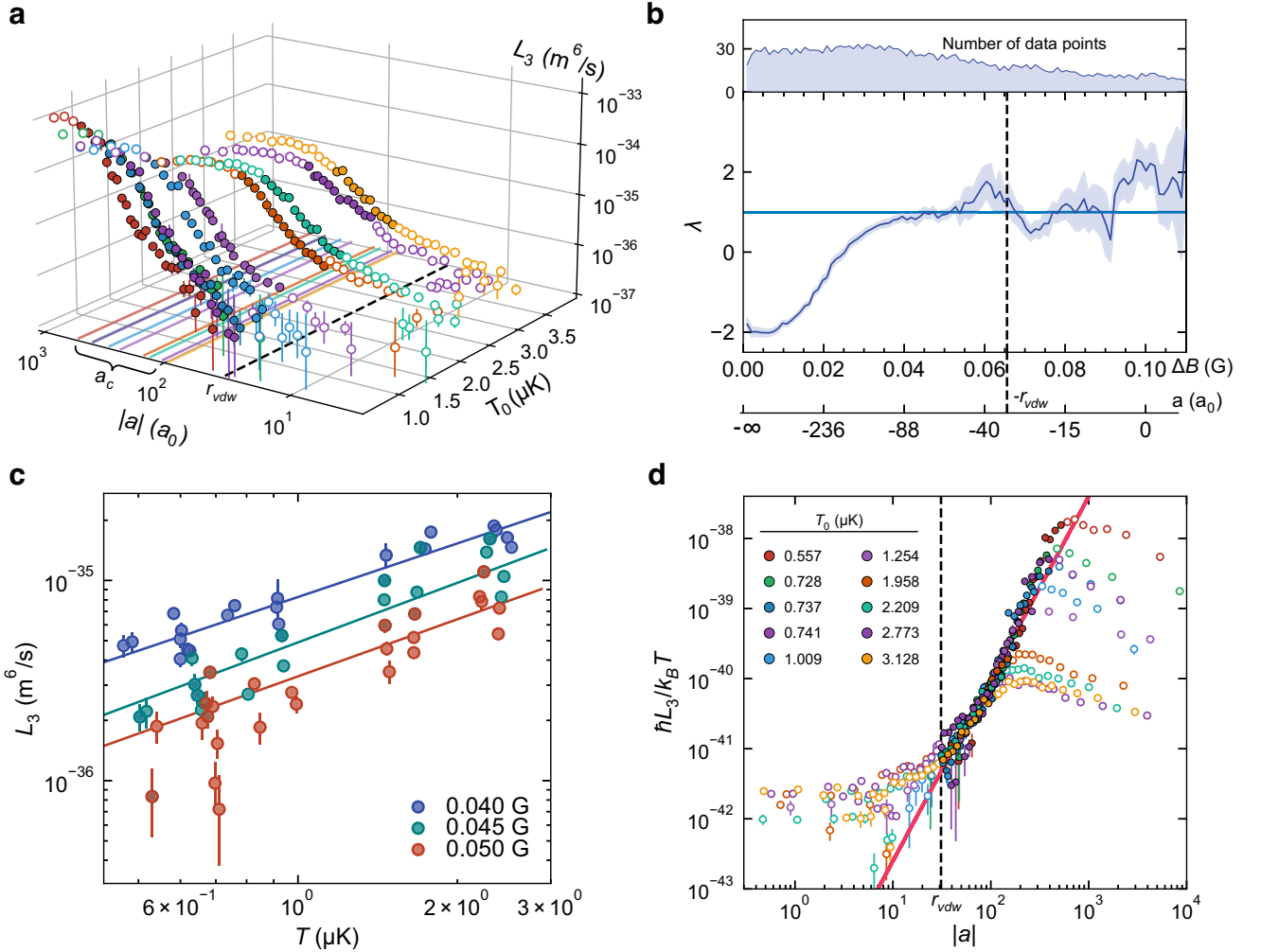}
		\caption{
			a) 3D plot of $L_3$ vs $|a|$ and initial temperature $T_0$. Colored solid lines denote the values of $a_c$ for different temperatures.
			b) plot of $\lambda$ vs $\Delta B$. Horizontal blue line indicates the $\lambda =1$. The light shadow shows the standard derivation of the fitting, and it becomes larger when fewer data points  are used.
			c) $L_3$ versus $T$ at $\Delta B$=0.04 G (blue), 0.045 G (cyan), and 0.05 G (red). Solid lines are their best-fitting.
			d) normalized three-body loss rate $\hbar L_3/k_B T$ vs $|a|$.
			Red line is the fitting of $L_3\propto |a|^{2.60(5)}$.
			In these figures, filled symbols are the data within an intermediate interacting regime, and the black dashed lines are the $r_{vdw}$ or the $\Delta B$ has a corresponding value of $|a|=r_{vdw}$.	
			Vertical error bars in these figures are the standard derivations. 		
		}\label{scalinglaw_a0}
	\end{center}
\end{figure*}

The two lowest hyperfine ground states mixed $^6$Li Fermi gas are loaded into a crossed-beam optical dipole trap. The gas temperature is lowered through a force evaporative cooling at a magnetic field of 300 G with a balanced spin. After cooling, the gas temperature is tuned by changing the final trap depth with an atom number of about $2\times10^5$ per spin. We also adjust the ratio between the trap depth and the gas temperature  so that the reduced temperatures $T/T_F$ are maintained around one and thus the gas can be treated as thermal. Here $T_F$ is the gas Fermi temperature.

The experimental procedure for the three-body atomic loss measurement is as follows.
First, the magnetic field is rapidly-swept over the narrow resonance to 570 G and held for a 100 ms to rethermalize the gas. Second, the magnetic field is precisely tuned in about 50 ms to the target field at which the three-body loss occurs~\cite{Chen2021PRA103.063311}. The target field is above the resonance $B_0=$\num{543.2704 +- 0.0016} G in the range of 10 to 100 mG, comparable to the resonance width $\Delta$=0.1 G. The stability of our magnetic field is controlled to 1.6 mG~\cite{Peng2022arXiv2107.07078}, and the gas is remains in the target field for a variable time of about 50 ms $\sim$ 450 ms. After the three-body interaction in the target field, the magnetic field is rapidly returned to 570 G within 5 ms for taking the time-of-flight absorption imaging.

We follow the procedure in Ref.~\cite{Li2018PRL120.193402} to attract the $L_3$ through the time-dependent atom number $N(t)$ with the three-body loss rate differential equation: $\dot{N(t)}=-L_3 V^{-2} N^3(t)$, where $V=(2\sqrt{3}\pi)^{3/2} \sigma_x \sigma_y \sigma_z$ and $\sigma_{x,y,z}$ denotes the Gaussian widths. In the experiments, $\sigma_{x,y,z}$ are also recorded with the absorption imaging.

\begin{figure}[htbp]
	\begin{center}
		\includegraphics[width=\columnwidth, angle=0]{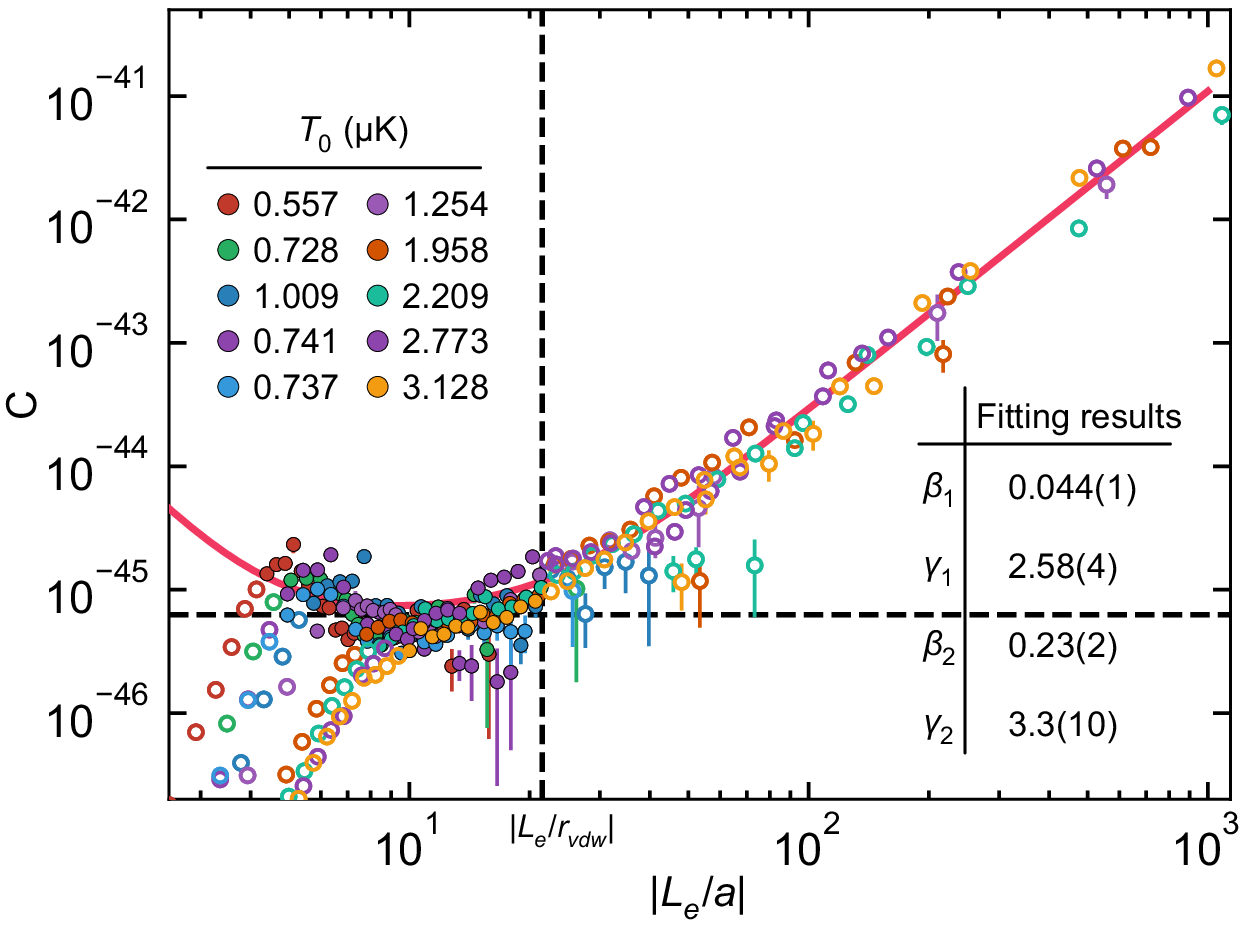}
		\caption{
		 Parameter $C$ versus $|L_e/a|$. Black line presents the baseline with an averaged value of $C_0=6.28\times 10^{-46}$ in the range of $5.5<L_e/|a|<20$. Filled symbols are the data in the intermediate interacting regime.		
		Red line displays the best-fitting of $C=C_0(1+ |\beta_1 L_e /a|^{\gamma_1}+|\beta_2 L_e /a|^{-\gamma_2})$ with a fixed $C_0$ and the fitted $\beta_1=0.044(1)$, $\gamma_1=2.58(4)$, $\beta_2=0.23(2)$, and $\gamma_2=3.3(10)$.	
		}\label{fig2all}
	\end{center}
\end{figure}

\begin{figure}[ht]
	\begin{center}
		\includegraphics[width=1\columnwidth, angle=0]{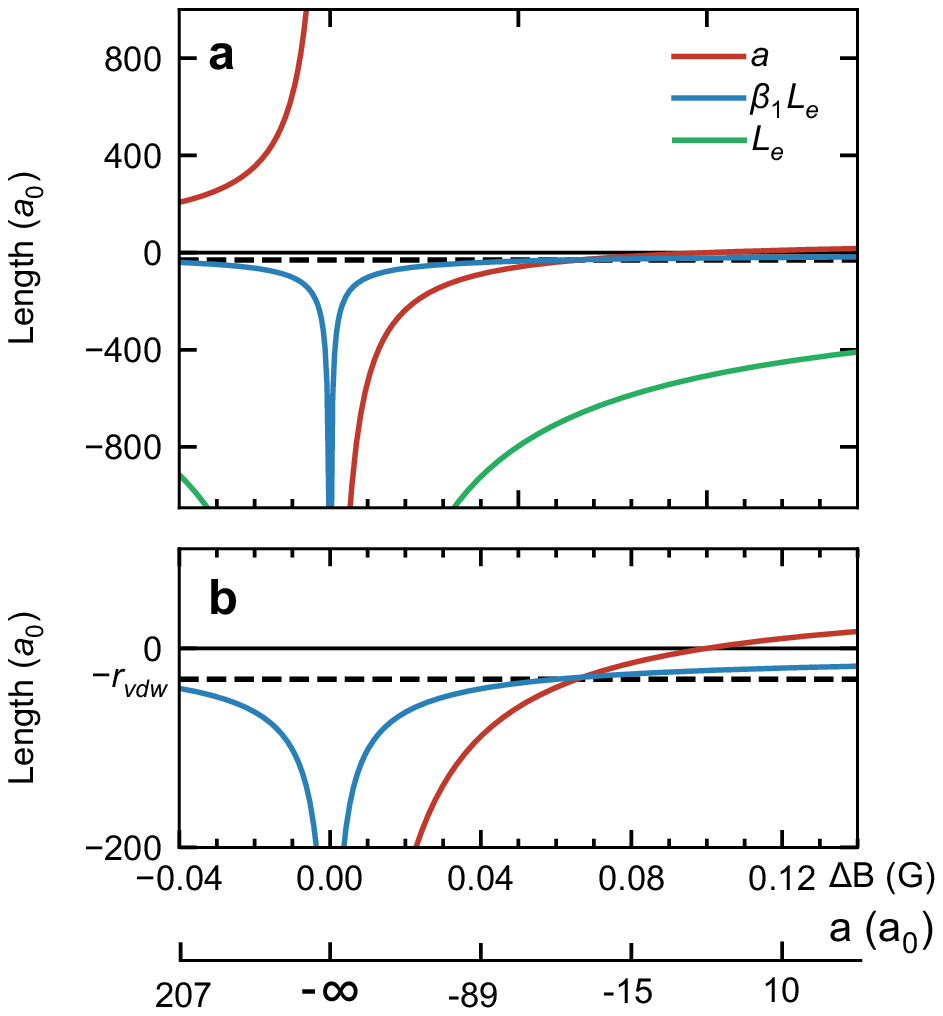}
		\caption{
			a) $a$ (red), $L_e$ (green), and $\beta_1 L_e$ (blue) versus $\Delta B$ near the narrow Feshbach resonance. b) A zoom-in figure on the vertical axis. Black dashed lines present $r_{vdw}$. $\beta_1 L_e$ shows a low change when $\Delta B>0.04$ G and roughly equals to $r_{vdw}$. $a$ cross with $\beta_1 L_e$ at $r_{vdw}$.
		}\label{Lengths}
	\end{center}
\end{figure}

We measure $L_3$ in a wide range of magnetic field detuning $\Delta B= B-B_0$ at various gas temperatures from 0.56 $\mu$K to 3.13 $\mu$K.
The results are shown in Fig.~\ref{scalinglaw_a0} a). Here, $\Delta B$ is converted to $|a|$ by a theory formula $a(B)=a_{bg}(1-\Delta/\Delta B)$ with a background scattering length $a_{bg}= 59$ $a_0$. In the thermal regime, $L_3(T)$ has a threshold behavior of $L_3\propto T$ given by the theoretical averaging of the three-body recombination rate $K_3(E)$ over the collision energy $E$ with a Maxwell-Boltzmann distribution.
Figure~\ref{scalinglaw_a0} b) shows the fit of the magnetic field dependence of the threshold behavior $L_3\propto T^{\lambda}$, where $\lambda$ is a free parameter. It is found that $\lambda$ equals to $-2$ at resonance, which agrees with the prediction~\cite{DIncao2018JPBAMOP51.043001}, and it starts to increase and reaches the value of one (blue horizontal line) until $\Delta B=$ 0.04 G. Below this value, the unitary behavior significantly affects the scaling. We verify this conclusion by refitting the data with a narrower temperature range in the Supplementary Materials (SM), which shows $\lambda$ equals to one in a range of 0.02 G$<\Delta B<$ 0.12 G.
Therefore, we conclude that the threshold behavior $L_3\propto T$ is not only valid in the intermediate interacting regime but also in the weakly interacting regime. Figure ~\ref{scalinglaw_a0} c) shows a plot of $L_3$ as a function of $T$ when $\Delta B=$ 0.04 G, 0.045 G, and 0.05 G, respectively. They show a monotonic increase over the temperature range we studied and can be well fitted by $\ln(L_3) = \lambda\ln(T) + b $ with $\lambda = 0.89(8), 0.99(14), 0.95(13)$.

We find that the unitary effect near the resonance limits the scaling behavior, requiring $|a|<a_c$.
We plot the values of $a_c$ for different temperatures as solid lines in Fig.~\ref{scalinglaw_a0}a). It can be seen that $a_c$ decreases with increasing $T_0$. So, when $T=8.7 \mu$K, $a_c=r_{vdw} \sim 31 a_0$ (black dashed line), indicating that the intermediate interacting regime disappears. To maintain this regime, the gas temperatures in this work are kept below 3 $\mu$K, which is lower than the temperature in the previous studies of three-body recombination near the narrow resonance~\cite{Li2018PRL120.193402, Hazlett2012PRL108.045304a}.
This proper temperature range is one of the key factors to implement the measurements presented in this work.

Next, we study the scattering length scaling law. As seen in Fig.~\ref{scalinglaw_a0}a), $L_3$ begins to increase linearly with $|a|$ when $|a|>r_{vdw}$ (black dashed line) and levels off after $|a|>a_c$ (colored solid lines) in the log-log plot.
In the intermediate regime $r_{vdw}<|a|<a_c$ (filled symbols), $L_3$ shows scaling law dependence of $|a|$. We normalize $L_3$ by $k_B T$ and plot it as a function of $|a|$ in Fig.~\ref{scalinglaw_a0}d).
Due to the three-body heating, the measured magnetic field dependence of the gas temperature is used to correct the data, which reduces some errors in the fits of Figs.~\ref{scalinglaw_a0}b)-d) (more details at SM).
Note that all data in the intermediate interacting regime collapse to a single linear line.
We fit the data in the intermediate interacting regime with
\begin{equation}\label{L3_eq1}
	L_3(a)=C\frac{k_B}{\hbar} {T}|a|^{\xi},
\end{equation}
and the best-fit results is $\xi=2.60(5)$ with $C=6.28(146)\times 10^{-46}$. This result is close to the prediction of the zero-range model for three-body recombination in two-component Fermi mixtures, $|a|^{2.455}$ in the BCS side of the Feshbach resonance rather than $|a|^6$ in the BEC side due to the three-body repulsive effective potential~\cite{DIncao2005PRL94.213201, DIncao2018JPBAMOP51.043001}.

Beyond the intermediate interacting regime, Figure~\ref{scalinglaw_a0} d) shows discrepancies from the $|a|^{2.60}$ scaling on both sides. To investigate these discrepancies, we introduce the effective scattering length $L_e$ and re-plot the data in Fig.~\ref{scalinglaw_a0}d) into the format of $C=\hbar L_3 /(k_B T |a|^{2.60})$ versus $|L_e/a|$, as shown in Fig.~\ref{fig2all}. 
Therefore, the discrepancies are significantly transferred into the deviations of parameter $C$ with a larger deviation appearing in the weakly interacting regime $|a|<r_{vdw}$ (vertical dashed line presents the $|L_e/r_{rdw}|$) and another smaller deviation in the stronger interacting regime $2<|L_e/a|<5.5$. 
It is known that $C$ remains constant in three-body recombination as long as the coupling strength between the quasi-bound and the deeper bound molecular states follows the same threshold and scaling behavior~\cite{Yoshida2018PRL120.133401}. Here, the deviations from $C$ are assumed as the consequence of finite-range effects and expressed with a correction of the $L_3$ in terms of $L_e/a$. The baseline of Fig.~\ref{fig2all} equals the fitted value of Fig.~\ref{scalinglaw_a0}d). The linearly increasing deviations in the log-log plot suggest power-law dependencies of $L_e/a$.
We fit the data with $C= C_0[1+|\beta_1 Le/a|^{\gamma_1}+|\beta_2 Le/a|^{-\gamma_2}]$ and obtain a best-fit result of $\beta_1 = 0.044(1)$, $\gamma_1=2.58(4)$, $\beta_2=0.23(2)$, and $\gamma_2=3.3(10)$. 

In the weakly interacting regime, because $|\beta_1 L_e/a|^{\gamma_1}$ dominates the scaling, the scaling law approximates as
\begin{equation}\label{L3_eq2}
	L_3(a, L_e)=C_0\frac{k_B T}{\hbar}(1+|\beta_1 L_e/a|^{\gamma_1})|a|^{\xi}.
\end{equation}
When $a\rightarrow 0$ and $\xi=2.60\approx\gamma_1=2.58$, the scaling law is further simplified as a single dependence on $L_e$ as $L_3=C_0 (k_B T/\hbar)|\beta_1 L_e|^{\gamma_1}$.
These scaling changes are also evident by analyzing the values of $|a|$, $L_e$, and $\beta_1 L_e$. As shown in Fig.~\ref{Lengths}, $|\beta_1 L_e|>|a|$ when $|a|<r_{vdw}$.
We noted that the same conclusion has also been found in the $s-$wave three-body boson systems~\cite{Shotan2014PRL113.053202}, where $L_e$ replaces $a$ as the length scale and coordinates the scaling behavior. Different from their experiments in the bosonic systems, a coefficient of $\beta_1$ is obtained from our fitting.
We should emphasize that $\beta_1$ plays an important role in our data analysis. It also makes $\beta_1 L_e$ roughly equals $r_{vdw}$, which is the single length scale when $a=0$.  

$L_e$ is energy-independent so that it maintains the threshold law of $L_3\propto T$ in the weakly interacting regime. And it will be different if other possible length scales, like $L^3_e k^2$, and $a(k)=a+L^3_e k^2$ ($k$ denotes the wavenumber of collision energy $E$), are used. 
Furthermore, since the measured $L_3$ does not vanish in the weakly interacting regime, even when $a=0$~\cite{Hazlett2012PRL108.045304a, Li2018PRL120.193402}, so neither $a$, nor $a(E)=a+L^3_e k^2$, nor $a+\beta_1 L_e$ can be used to describe the scaling due to their zero-crossing in the weakly interacting regime when $T\sim1\mu$K (more details at SM).
Therefore, we suggest that $L_e$ is the universal length at weakly interacting for the presented work. Further tests using two-component $s-$wave boson systems and other fermionic systems could be implemented to test this suggestion.

In the stronger interacting regime, the $|\beta_1 L_e|^{\gamma_1}$ can be ignored, and the scaling becomes
\begin{equation}\label{L3_eq3}
	L_3(a, L_e)=C_0\frac{k_BT}{\hbar} (1+|\beta_2 \frac{L_e}{a}|^{-\gamma_2})|a|^{\xi}.
\end{equation}
As shown in Eq.~\ref{L3_eq3}, the finite-range correction of $|\beta_2 L_e/a|^{-\gamma_2}$ accomplishes the original $|a|^{2.60}$ scaling, when $a\sim a_c$. So, in Eq~\ref{L3_eq3}, there are two terms related to $|a|$, which are the $|a|^{2.60(5)}$ scaling and the $|a|^{3.3(10)+2.60(5)=5.9(15)}$ scaling. This is different from the scaling in the weakly interacting case. 
The experiment in the spin-polarized gas of $^6$Li atoms near a $p-$wave Feshbach resonant recently also supports two scaling behaviors ~\cite{Yoshida2018PRL120.133401}, where the finite-range correction increases as interaction increases.
Both our $s-$wave experiment and their $p-$wave experiment show that finite-range correction of the three-body recombination has yet to be understood, expecting more experimental and theoretical studies in the future.

In summary, we have experimentally investigated the scaling law of the three-body loss rate of a two-component $^6$Li Fermi gas near its narrow $s-$wave Feshbach resonance, in which both the threshold behavior due to the unitary limit and the finite range effect are characterized.
We confirm that the threshold behavior of $L_3\propto T$ shown for the broad resonance is also not only valid for the narrow Feshbach resonance but also can be extended to the weakly interacting regime. In the intermediate interacting strength regime, we have identified a scaling law of $L_3\propto |a|^{2.60}$, which is close to the theoretical prediction of $|a|^{2.455}$, but with some deviation near the threshold regime around $a_c$. 
In the weakly interacting regime, we observe a large finite-range correction and obtain a scaling of $L_3\propto L^{2.58}_e$ with an effective scattering length $L_e$.
Additionally, we also observe another finite-range correction in the stronger interacting region, where the original scaling of $|a|^{2.60}$ always accompanies a smaller correction of $\propto |a|^{5.9}/|{L_e}|^{3.3}$. 

Since the unitary behavior is found to be dominant and changes the scaling law completely when $|a|>a_c$, we believe that Ref.~\cite{Du2009PRL102.250402} shows a $|a|^{\num{0.79 +- 0.14}}$ scaling near the broad Feshbach resonance of the two-component $^6$Li-Fermi gas, which could be a consequence of the unitary effect~\cite{unitaryBFR}. To verify this conjecture, we fit our data with a broader range of interaction strengths using the data from $r_{vdw} < |a|$ in Fig.~\ref{scalinglaw_a0}d) and obtain a scaling of $\propto |a|^{\num{0.73 +- 0.08}}$.
Thus, we believe that the large background scattering length near the broad Feshbach resonance messes up the original scaling of the three-body recombination.

\textit{Acknowledgments}
We thank Dr. Zhenhua Yu and Dr. Yangqian Yan for the helpful discussions. This work receives support from the National Natural Science Foundation of China under Grant No.12174458, 11804406, and 11774436,
and from the Key-Area Research and Development Program of Guangdong Province under Grant No. 2019B030330001. The Fundamental Research Funds for the Central Universities, Sun Yat-sen University under Grant No. 2021qntd28. LL thank the supports from Guangdong Province Youth Talent Program under Grant No.2017GC010656.

	
	
%

\newpage
\newpage

\section{Supplementary Materials}

\setcounter{equation}{0}
\setcounter{subsection}{0}
\setcounter{figure}{0}
\renewcommand{\theequation}{S.\arabic{equation}}
\renewcommand{\thesubsection}{S.\arabic{subsection}}
\renewcommand{\thefigure}{S.\arabic{figure}}	
\subsection*{S1: Corrections of the temperature fluctuation}
\begin{figure}[htbp]
	\begin{center}
		\includegraphics[width=\columnwidth, angle=0]{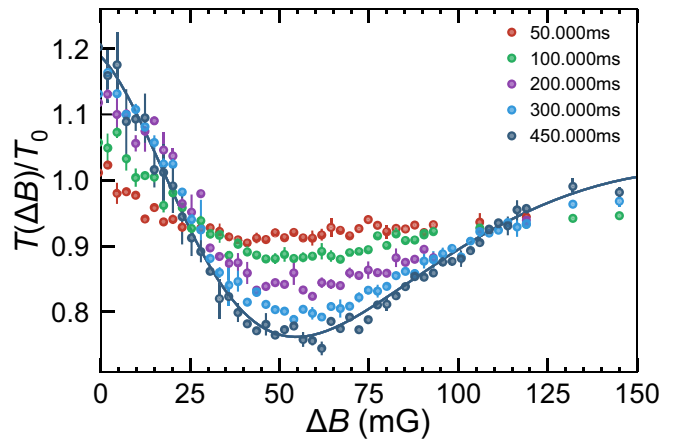}
		\caption{A typical $T/T_0$ versus $\Delta B$ for $T_0=3.13\mathrm{\mu K}$. The different color present the different holding time durations. The temperature drops is due to three-body cooling of our work~\cite{Peng2022arXiv2107.07078(SM)}. The solid line is a smoothing spline of the temperatures. }\label{SM:FigT}
	\end{center}
\end{figure}

We observe a magnetic field dependent temperature fluctuation $T(\Delta B)/T_0$, which is due to three-body heating or cooling~\cite{Peng2022arXiv2107.07078(SM)}, where $T_0$ stands for the initial temperature measured at a far-off resonance of a 570 G magnetic field.
Figure~\ref{SM:FigT} shows a typical $T$ as a function of the magnetic field detuning $\Delta B$. It can be seen that the temperature fluctuation is less than $20\%$ in the most magnetic fields. 

In Fig.1 b and c, the $T^{\lambda}$ is obtained by fitting the $L_3$ at different temperatures for a given $\Delta B$, and we use correction $T$ instead of the initial temperature $T_0$. 
In Fig.1 d, the correction $T$ is also used to obtain the temperature normalized three-body atomic loss rate $\hbar L_3/k_B T$.

\subsection*{S2: The range of the intermediate interacting regime}
The scattering length scaling law of the $L_3$ holds in the regime of indeterminate interacting, where $|a|$ lies between the lower limit of $r_{vdw}$ and the upper limit of $a_c$. Here, $a_c$ characterizes the condition of the unitary regime that can be determined by the ratio between the thermal energy and the quasi-molecule binding energy when $a<0$. When the binding energy is smaller than the thermal energy, we treat the system in the unitary regime.

\begin{figure}[htbp]
	\begin{center}
		\includegraphics[width=\columnwidth, angle=0]{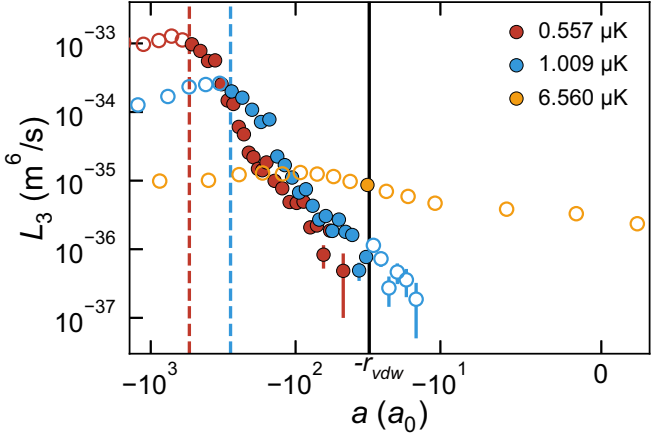}
		\caption{$L_3$ as a function of $a$ at various temperatures of 0.557 $\mu$K (red), 1.009 $\mu$K (blue) and 6.560 $\mu$K (orange), respectively. Black solid line is $r_{vdw}=31a_0$. Colored dashed lines present their $a_c$. The filled symbols between $r_{rdw}<|a|<a_c$ are the data in the intermediate interacting regime, where we use to obtain the scaling law. For the orange data, the range of intermidiante interacting regime almost vanishes. }\label{SM:Fig_Rawdata}
	\end{center}
\end{figure}

In a narrow Feshbach resonance, the two-body $s-$wave collision amplitude $f$ has an effective-range factor $r_{eff}$ as
\begin{equation}
	f=-\frac{1}{a^{-1}-r_{eff}k^2/2+ik}.
\end{equation}
At the pole of $f$, we obtain the binding energy of the molecule $E_b=\hbar^2/(2\mu a'^2)$, where $a'=r_{eff}/(-\sqrt{-1+2r_{eff}/a}+i)$ and $\mu$ is the reduced mass of two-body system.
We approach $E_b=\hbar^2/(\mu r_{eff}a)$ when $r_{eff}\gg a$. 
Since the gas temperatures are around $T/T_F\sim1$, we use $3k_BT$ to the averaged thermal energy of a trapped atom. 
When $E_b=3k_BT$, we obtain $a_c=2\hbar^2/(3m|r_{eff}|k_BT)$.

Figure~\ref{SM:Fig_Rawdata} shows $L_3$ at different temperatures. For the red and blue data (low temperatures), $L_3$ decreases when $a$ approach the resonance point, we label the $a_c$ as the red and blue dashed lines. It can be seen that the turning point is around $a_c$. But for orange data (high temperature), the decreasing is not obvious. These data show $a_c$ decrease as $T$ increase.   
When $a_c=r_{vdw}$, $T=8.7 \mu$K, then the intermediate interacting regime vanishes. This indicate that, to obtain the scaling law in the intermediate interacting regime, we should use the gas temperature $T$ well below 8.7 $\mu$K.

\subsection*{S3: Threshold behavior near the resonant}
In the main text, we state that the fitted $\lambda$ in Fig.1 b) deviates from one when the magnetic field is closed to the resonant. In this section, we provide more details. 

\begin{figure}[htbp]
	\begin{center}
		\includegraphics[width=\columnwidth, angle=0]{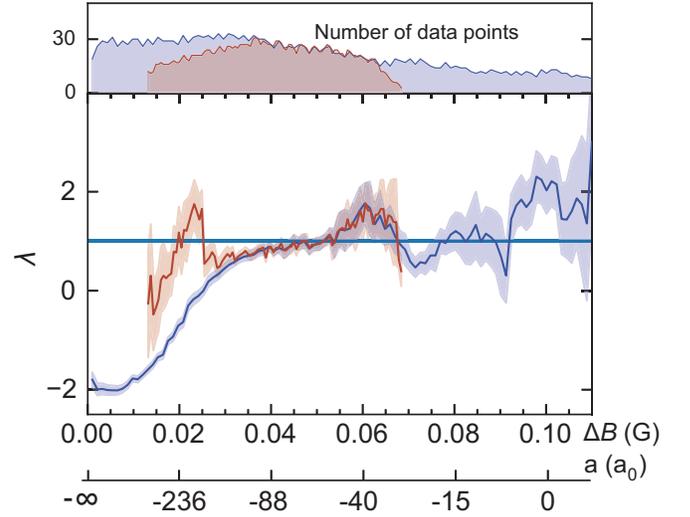}
		\caption{The fitted $\lambda$ with all temperature data (blue curve) and chosen low temperature data (red curve). Top panel displays the number of temperature points used in two fits.  
			The horizontal line presents $\lambda=1$, and the light shadows are the standard derivation of the fittings.}\label{SM:thresholdLaw_com}
	\end{center}
\end{figure}

We obtain the $\lambda$ by fitting the $L_3$ with a power law of $T^{\lambda}$ at a given magnetic field detuning $\Delta B$. The data in $\Delta B\pm$ 3.5 mG are binned to the data of $\Delta B$ because of the magnetic field fluctuation. The blue curve in Fig.~\ref{SM:thresholdLaw_com} shows the result of $\lambda$ from 0.56 $\mathrm{\mu K}$ to 3.13 $\mathrm{\mu K}$. Then, we choose the data in a narrow range of temperature, $T < 2\hbar^2/(3mk_Br_{eff}a)$, to draw the red curve. The reason is that when  $T> 2\hbar^2/(3mk_Br_{eff}a)$, the data are in the unitary regime as we descried in the main text. 
Figure S.3 shows that by redrawing the data in a narrower range (red curve), $\lambda$ returns to one in the regime of 0.015 G $< \Delta B < $0.04 G. Thus, the divergence of the blue curve in the above regime is because some high temperature data are beyond the intermediate interacting regime.  
When $\Delta B<$0.015 G, the temperature range of the intermediate interacting regime decrease, meaning we need very low temperature data to draw the red curve. Limited by our experiment, it is difficult to generate a very low temperature data to satisfy the condition of the intermediate interacting regime. So, the red curve stops around 0.015 G.

\subsection*{S4: Energy dependence of scattering length: effective scattering length $L_e$}
The two-body scattering length of the narrow \textit{s}-wave Feshbach resonance of $^6$Li can be approximated as
\begin{equation}
	a = a_{bg}(1-\frac{\Delta}{\Delta B}),
\end{equation}
where resonance width $\Delta=0.1$ G, background scattering length $a_{bg}$=59 $a_0$, and  resonance center $B_0=543.2704$ G.
In a thermal gas, the collision energy $E=\hbar^2 k^2/2\mu$, and the low-energy effective expansion of $a$ is 
\begin{equation}
	a(E)=a+\frac{1}{2}r_{eff}a^2k^2.
\end{equation}

\begin{figure}[htbp]
	\begin{center}
		\includegraphics[width=0.9\columnwidth, angle=0]{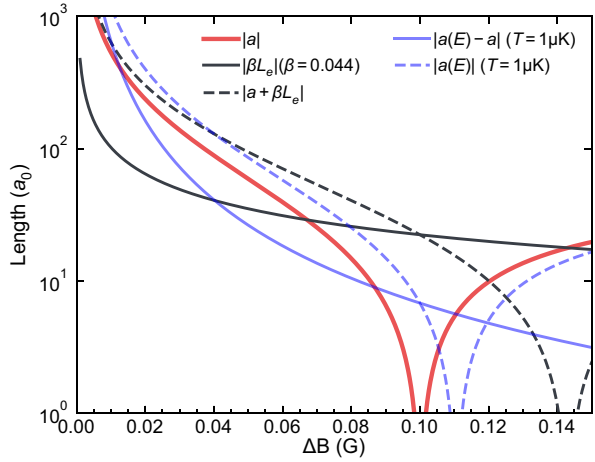}
		\caption{Several typical effective scattering lengths versus $\Delta B$. $k_T$ is the thermal wave number. $a$ is the two-body $s-$wave scattering length. $\beta L_e$ is the scaled used in our experiment and Ref.18 in the main text.  $a(E)$, $(a+\beta L_e)$ and $[a(E)-a]$ are the other possible candidates for the length scale.
		}\label{SM:lengthSum}
	\end{center}
\end{figure}

Here, the value of the effective collision length $L_e=(r_{eff}a^2/2)^{1/3}$ is taken from Ref.~\cite{Blackley2014PRA89.042701(SM)}, where it is obtained by a coupled channel calculation.
For more convenient, we use its parabolic approximation given by Ref.~\cite{Blackley2014PRA89.042701(SM)} for $^6$Li narrow $s-$wave Feshbach resonance as
\begin{equation}
	r_{eff}a^2=v+r_0(a-a_{ext})^2,
\end{equation}
to calculate the value of $L_e$,
where $v=-4.9\times 10^6 a^3_0$, $r_0=-71000 a_0$, and $a_{ext}=60 a_0$. 
The effective collision length is important in studying the finite-range corrections of the scaling law. Especially, the measured $L_3$ does not vanish as $a$ at the zero-crossing. 

Figure~\ref{SM:lengthSum} plots several candidates for the new length scales. It can be seen that $L_e$ and its reduced value $\beta L_e$ keeps finite at $a=0$ and other length scales show a zero-crossing near $a=0$.  This indicates that for $a=0$, $L_e$ and $\beta L_e$ could be a nonzero length scale to describe the behavior of the systems.

\subsection*{S5: Experimental details}
The experiment parameters, contain trapping frequencies $\omega_{x, y, z}$, initial atom number $N_0$, Fermi temperature $T_F$, and reduced temperature $T_0/T_F$, of our data at different temperatures are listed in Tab.~\ref{tab:essential_parameters}. 

\begin{table}[htbp]
	\caption{Summary of $\omega_{x, y, z}$, $N_0$, $T_F$ and $T_0/T_F$ at various temperatures $T_0$.}
	\label{tab:essential_parameters}
	\renewcommand{\arraystretch}{1.5}
	\resizebox{1\columnwidth}{!}{%
		\begin{tabular}{ccccccc}
			\hline
			$T_0 (\mathrm{\mu K})$ & $\omega_x (\mathrm{Hz})$ & $\omega_y (\mathrm{Hz})$ & $\omega_z (\mathrm{Hz})$ & $N_0$ & $T_F$ ($\mathrm{\mu K}$) & $T_0/T_F$ \\ \hline
			0.557 & 2281.9 & 2282.9 & 203.1 & 1.25$\times 10^5$  & 0.71 & 0.79 \\
			0.728 & 2791.9 & 2797.5 & 264.7 & 1.60$\times 10^5$   & 0.96 & 0.76 \\
			0.737 & 2629.2 & 2638.9 & 261.4 & 2.10$\times 10^5$   & 1.01 & 0.73 \\
			0.741 & 2459.4 & 2468.5 & 244.5 & 2.02$\times 10^5$  & 0.93 & 0.80 \\
			1.009 & 3221.4 & 3231.5 & 314.1 & 1.85$\times 10^5$  & 1.17 & 0.86 \\
			1.254 & 3718.3 & 3731.9 & 369.7 & 3.00$\times 10^5$     & 1.60 & 0.78 \\
			1.958 & 4157.2 & 4172.4 & 413.4 & 2.92$\times 10^5$  & 1.78 & 1.10 \\
			2.210 & 4554.0 & 4570.7 & 452.8 & 3.05$\times 10^5$  & 1.97 & 1.12 \\
			2.773 & 5258.5 & 5277.8 & 522.9 & 3.00$\times 10^5$     & 2.27 & 1.22 \\
			3.128 & 4918.8 & 4936.9 & 489.1 & 3.22$\times 10^5$  & 2.17 & 1.44 \\
			\hline
		\end{tabular}%
	}
\end{table}

\subsection*{S7: Raw Data}
\newpage
\renewcommand{\arraystretch}{1}
\begin{longtable}{@{}p{1.5cm}p{1.5cm}p{2.2cm}p{2.2cm}@{}} 
	\caption{Raw data in the main text.}
	\label{tab:Raw_data}\\
	\toprule
	$T_0$($\mu$K)      & $\Delta B$(mGs) & $L_3$($\times 10^{-36}\mathrm{m^6}$/s) & $\sigma (L_3)$($\times 10^{-36}\mathrm{m^6}$/s) \\* \midrule
	\endfirsthead
	\endhead
	\bottomrule
	\endfoot
	\endlastfoot
	0.557     & 1.09            & 426.198                 & 56.563                         \\
	0.557     & 2.39            & 867.059                 & 37.868                         \\
	0.557     & 3.69            & 1029.051                & 41.222                         \\
	0.557     & 4.99            & 970.742                 & 18.615                         \\
	0.557     & 6.29            & 1101.383                & 53.897                         \\
	0.557     & 7.59            & 1277.342                & 70.087                         \\
	0.557     & 8.89            & 1123.803                & 81.089                         \\
	0.557     & 10.19           & 971.887                 & 69.768                         \\
	0.557     & 11.49           & 777.339                 & 76.364                         \\
	0.557     & 12.79           & 554.508                 & 52.861                         \\
	0.557     & 14.10           & 565.887                 & 83.862                         \\
	0.557     & 15.40           & 254.082                 & 25.367                         \\
	0.557     & 16.70           & 147.089                 & 11.723                         \\
	0.557     & 18.00           & 130.951                 & 12.178                         \\
	0.557     & 19.30           & 60.697                  & 4.435                          \\
	0.557     & 20.60           & 47.342                  & 3.742                          \\
	0.557     & 21.90           & 25.530                  & 1.613                          \\
	0.557     & 23.20           & 21.920                  & 1.238                          \\
	0.557     & 24.50           & 14.833                  & 0.788                          \\
	0.557     & 25.80           & 14.148                  & 0.899                          \\
	0.557     & 27.10           & 18.405                  & 1.585                          \\
	0.557     & 29.71           & 9.920                   & 0.562                          \\
	0.557     & 32.31           & 7.728                   & 0.753                          \\
	0.557     & 34.91           & 4.878                   & 0.522                          \\
	0.557     & 37.51           & 4.729                   & 0.600                          \\
	0.557     & 40.11           & 4.936                   & 0.562                          \\
	0.557     & 42.72           & 2.082                   & 0.324                          \\
	0.557     & 45.32           & 2.218                   & 0.383                          \\
	0.557     & 47.92           & 0.833                   & 0.311                          \\
	0.557     & 50.52           & 1.868                   & 0.338                          \\
	0.557     & 55.73           & 0.484                   & 0.384                          \\
	0.728     & 0.69            & 187.000                 & 6.870                          \\
	0.728     & 3.29            & 285.000                 & 14.000                         \\
	0.728     & 5.89            & 422.000                 & 16.200                         \\
	0.728     & 8.49            & 539.000                 & 16.500                         \\
	0.728     & 11.09           & 579.000                 & 32.200                         \\
	0.728     & 13.70           & 434.000                 & 17.100                         \\
	0.728     & 16.30           & 249.000                 & 16.400                         \\
	0.728     & 18.90           & 148.000                 & 8.670                          \\
	0.728     & 21.50           & 80.200                  & 6.160                          \\
	0.728     & 22.80           & 59.800                  & 4.740                          \\
	0.728     & 24.10           & 44.500                  & 3.470                          \\
	0.728     & 25.40           & 28.400                  & 1.650                          \\
	0.728     & 26.70           & 19.900                  & 0.957                          \\
	0.728     & 28.01           & 13.800                  & 0.760                          \\
	0.728     & 29.31           & 13.100                  & 0.507                          \\
	0.728     & 30.61           & 10.100                  & 0.664                          \\
	0.728     & 31.91           & 7.390                   & 0.553                          \\
	0.728     & 33.21           & 8.910                   & 0.524                          \\
	0.728     & 34.51           & 6.810                   & 0.560                          \\
	0.728     & 35.81           & 7.050                   & 0.593                          \\
	0.728     & 37.11           & 6.840                   & 0.368                          \\
	0.728     & 39.71           & 5.100                   & 0.477                          \\
	0.728     & 44.92           & 4.090                   & 0.357                          \\
	0.728     & 50.12           & 1.940                   & 0.344                          \\
	0.728     & 52.72           & 2.430                   & 0.370                          \\
	0.728     & 55.33           & 0.675                   & 0.518                          \\
	0.728     & 70.94           & 0.384                   & 0.317                          \\
	0.737     & 0.12            & 67.071                  & 4.394                          \\
	0.737     & 2.72            & 135.837                 & 3.913                          \\
	0.737     & 5.33            & 218.989                 & 12.317                         \\
	0.737     & 7.93            & 278.236                 & 9.343                          \\
	0.737     & 10.53           & 343.895                 & 8.396                          \\
	0.737     & 13.13           & 277.264                 & 10.148                         \\
	0.737     & 15.73           & 221.894                 & 6.245                          \\
	0.737     & 18.33           & 145.798                 & 6.541                          \\
	0.737     & 20.94           & 83.524                  & 9.006                          \\
	0.737     & 23.54           & 44.938                  & 1.709                          \\
	0.737     & 26.14           & 20.345                  & 1.230                          \\
	0.737     & 28.74           & 11.739                  & 0.455                          \\
	0.737     & 31.34           & 8.195                   & 0.324                          \\
	0.737     & 33.95           & 7.142                   & 0.575                          \\
	0.737     & 36.55           & 4.052                   & 0.343                          \\
	0.737     & 39.15           & 4.527                   & 0.348                          \\
	0.737     & 41.75           & 3.021                   & 0.400                          \\
	0.737     & 44.35           & 2.258                   & 0.260                          \\
	0.737     & 46.96           & 2.089                   & 0.267                          \\
	0.737     & 49.56           & 2.347                   & 0.278                          \\
	0.737     & 52.16           & 1.533                   & 0.253                          \\
	0.737     & 54.76           & 1.025                   & 0.228                          \\
	0.737     & 57.36           & 0.901                   & 0.219                          \\
	0.737     & 59.96           & 0.616                   & 0.308                          \\
	0.737     & 62.57           & 0.470                   & 0.158                          \\
	0.737     & 70.37           & 0.400                   & 0.259                          \\
	0.741     & 0.12            & 93.953                  & 7.041                          \\
	0.741     & 2.72            & 142.917                 & 5.608                          \\
	0.741     & 5.33            & 211.228                 & 7.591                          \\
	0.741     & 7.93            & 287.682                 & 9.142                          \\
	0.741     & 10.53           & 426.035                 & 12.220                         \\
	0.741     & 13.13           & 395.941                 & 6.968                          \\
	0.741     & 15.73           & 328.559                 & 13.895                         \\
	0.741     & 18.33           & 194.277                 & 4.761                          \\
	0.741     & 20.94           & 162.876                 & 7.737                          \\
	0.741     & 23.54           & 60.176                  & 3.071                          \\
	0.741     & 26.14           & 31.264                  & 1.737                          \\
	0.741     & 28.74           & 15.701                  & 0.879                          \\
	0.741     & 31.34           & 14.801                  & 2.276                          \\
	0.741     & 33.95           & 7.775                   & 0.429                          \\
	0.741     & 36.55           & 5.630                   & 0.494                          \\
	0.741     & 39.15           & 4.450                   & 0.238                          \\
	0.741     & 41.75           & 2.670                   & 0.281                          \\
	0.741     & 44.35           & 2.447                   & 0.314                          \\
	0.741     & 46.96           & 3.476                   & 0.242                          \\
	0.741     & 49.56           & 0.971                   & 0.270                          \\
	0.741     & 52.16           & 0.720                   & 0.344                          \\
	0.741     & 57.36           & 0.319                   & 0.273                          \\
	0.741     & 59.96           & 0.290                   & 0.222                          \\
	0.741     & 67.77           & 1.059                   & 0.183                          \\
	1.009     & 1.99            & 29.756                  & 4.964                          \\
	1.009     & 4.59            & 126.854                 & 3.290                          \\
	1.009     & 7.19            & 168.431                 & 6.678                          \\
	1.009     & 9.79            & 232.641                 & 5.401                          \\
	1.009     & 12.39           & 251.577                 & 10.508                         \\
	1.009     & 15.00           & 262.024                 & 15.960                         \\
	1.009     & 17.60           & 199.441                 & 9.937                          \\
	1.009     & 20.20           & 161.921                 & 9.055                          \\
	1.009     & 22.80           & 107.915                 & 8.138                          \\
	1.009     & 25.40           & 71.585                  & 2.511                          \\
	1.009     & 28.01           & 77.943                  & 4.167                          \\
	1.009     & 30.61           & 22.496                  & 1.250                          \\
	1.009     & 33.21           & 16.817                  & 0.824                          \\
	1.009     & 35.81           & 11.085                  & 0.239                          \\
	1.009     & 38.41           & 6.703                   & 0.232                          \\
	1.009     & 41.02           & 7.489                   & 0.368                          \\
	1.009     & 43.62           & 4.297                   & 0.336                          \\
	1.009     & 46.22           & 2.702                   & 0.169                          \\
	1.009     & 48.82           & 3.047                   & 0.143                          \\
	1.009     & 51.42           & 1.848                   & 0.329                          \\
	1.009     & 54.02           & 2.706                   & 0.242                          \\
	1.009     & 56.63           & 1.794                   & 0.269                          \\
	1.009     & 59.23           & 1.614                   & 0.212                          \\
	1.009     & 61.83           & 0.491                   & 0.146                          \\
	1.009     & 64.43           & 0.771                   & 0.146                          \\
	1.009     & 67.03           & 1.139                   & 0.193                          \\
	1.009     & 69.64           & 0.725                   & 0.174                          \\
	1.009     & 72.24           & 0.272                   & 0.128                          \\
	1.009     & 74.84           & 0.466                   & 0.156                          \\
	1.009     & 77.44           & 0.358                   & 0.158                          \\
	1.009     & 80.04           & 0.187                   & 0.137                          \\
	1.254     & 0.07            & 42.079                  & 2.809                          \\
	1.254     & 1.37            & 65.807                  & 5.209                          \\
	1.254     & 2.68            & 75.198                  & 4.819                          \\
	1.254     & 7.88            & 119.685                 & 7.309                          \\
	1.254     & 13.08           & 130.280                 & 3.909                          \\
	1.254     & 18.29           & 120.530                 & 6.543                          \\
	1.254     & 23.49           & 76.989                  & 3.499                          \\
	1.254     & 26.09           & 59.935                  & 11.637                         \\
	1.254     & 27.39           & 42.476                  & 2.640                          \\
	1.254     & 28.69           & 38.028                  & 2.651                          \\
	1.254     & 30.00           & 29.562                  & 1.877                          \\
	1.254     & 32.60           & 20.758                  & 1.115                          \\
	1.254     & 33.90           & 18.786                  & 1.459                          \\
	1.254     & 35.20           & 14.613                  & 1.150                          \\
	1.254     & 37.80           & 7.363                   & 0.521                          \\
	1.254     & 39.10           & 8.142                   & 1.914                          \\
	1.254     & 40.40           & 6.058                   & 0.375                          \\
	1.254     & 43.00           & 5.306                   & 0.248                          \\
	1.254     & 44.31           & 3.744                   & 0.243                          \\
	1.254     & 49.51           & 2.758                   & 0.098                          \\
	1.254     & 52.11           & 2.410                   & 0.234                          \\
	1.254     & 54.71           & 1.797                   & 0.144                          \\
	1.254     & 59.92           & 1.628                   & 0.148                          \\
	1.254     & 65.12           & 1.501                   & 0.122                          \\
	1.254     & 70.32           & 1.073                   & 0.062                          \\
	1.254     & 75.53           & 0.831                   & 0.104                          \\
	1.254     & 78.13           & 0.471                   & 0.087                          \\
	1.254     & 80.73           & 0.402                   & 0.126                          \\
	1.254     & 85.93           & 0.485                   & 0.068                          \\
	1.254     & 101.55          & 0.243                   & 0.057                          \\
	1.254     & 104.15          & 0.269                   & 0.109                          \\
	1.254     & 117.16          & 0.205                   & 0.106                          \\
	1.958     & 2.54            & 24.521                  & 1.453                          \\
	1.958     & 5.14            & 30.675                  & 1.498                          \\
	1.958     & 7.74            & 33.620                  & 1.084                          \\
	1.958     & 10.34           & 41.230                  & 1.354                          \\
	1.958     & 12.95           & 49.369                  & 2.159                          \\
	1.958     & 15.55           & 50.008                  & 1.182                          \\
	1.958     & 18.15           & 54.650                  & 1.050                          \\
	1.958     & 20.75           & 52.578                  & 0.625                          \\
	1.958     & 23.35           & 51.163                  & 1.116                          \\
	1.958     & 25.96           & 39.892                  & 1.794                          \\
	1.958     & 28.56           & 40.001                  & 0.916                          \\
	1.958     & 31.16           & 31.911                  & 1.079                          \\
	1.958     & 33.76           & 25.576                  & 1.077                          \\
	1.958     & 36.36           & 19.103                  & 1.190                          \\
	1.958     & 38.96           & 13.369                  & 1.910                          \\
	1.958     & 41.57           & 9.997                   & 0.621                          \\
	1.958     & 44.17           & 8.005                   & 0.508                          \\
	1.958     & 46.77           & 5.963                   & 0.479                          \\
	1.958     & 49.37           & 4.556                   & 0.353                          \\
	1.958     & 51.97           & 3.498                   & 0.464                          \\
	1.958     & 54.58           & 3.091                   & 0.189                          \\
	1.958     & 57.18           & 2.328                   & 0.164                          \\
	1.958     & 59.78           & 2.269                   & 0.257                          \\
	1.958     & 62.38           & 1.868                   & 0.131                          \\
	1.958     & 64.98           & 1.522                   & 0.141                          \\
	1.958     & 67.58           & 2.041                   & 0.087                          \\
	1.958     & 70.19           & 1.691                   & 0.056                          \\
	1.958     & 72.79           & 1.577                   & 0.073                          \\
	1.958     & 75.39           & 1.210                   & 0.090                          \\
	1.958     & 77.99           & 1.057                   & 0.073                          \\
	1.958     & 80.59           & 1.321                   & 0.063                          \\
	1.958     & 83.20           & 1.196                   & 0.077                          \\
	1.958     & 85.80           & 0.957                   & 0.064                          \\
	1.958     & 88.40           & 1.023                   & 0.044                          \\
	1.958     & 91.00           & 0.390                   & 0.067                          \\
	1.958     & 93.60           & 0.642                   & 0.047                          \\
	1.958     & 96.20           & 0.534                   & 0.067                          \\
	1.958     & 98.81           & 0.398                   & 0.042                          \\
	1.958     & 101.41          & 0.567                   & 0.059                          \\
	1.958     & 104.01          & 0.175                   & 0.052                          \\
	1.958     & 117.02          & 0.082                   & 0.047                          \\
	2.210     & 0.34            & 14.548                  & 0.585                          \\
	2.210     & 2.95            & 12.375                  & 1.115                          \\
	2.210     & 5.55            & 24.797                  & 0.456                          \\
	2.210     & 8.15            & 28.049                  & 0.351                          \\
	2.210     & 10.75           & 32.432                  & 0.543                          \\
	2.210     & 13.35           & 34.824                  & 0.362                          \\
	2.210     & 15.96           & 31.714                  & 0.566                          \\
	2.210     & 18.56           & 38.616                  & 0.619                          \\
	2.210     & 21.16           & 40.798                  & 0.664                          \\
	2.210     & 23.76           & 37.786                  & 0.657                          \\
	2.210     & 26.36           & 35.486                  & 0.821                          \\
	2.210     & 28.96           & 34.065                  & 1.062                          \\
	2.210     & 31.57           & 26.354                  & 0.472                          \\
	2.210     & 34.17           & 21.342                  & 0.479                          \\
	2.210     & 36.77           & 17.451                  & 0.312                          \\
	2.210     & 39.37           & 14.388                  & 0.304                          \\
	2.210     & 41.97           & 14.592                  & 0.652                          \\
	2.210     & 44.58           & 8.739                   & 0.167                          \\
	2.210     & 47.18           & 6.804                   & 0.188                          \\
	2.210     & 49.78           & 5.191                   & 0.098                          \\
	2.210     & 52.38           & 4.360                   & 0.137                          \\
	2.210     & 54.98           & 4.565                   & 0.199                          \\
	2.210     & 57.58           & 2.782                   & 0.065                          \\
	2.210     & 60.19           & 2.203                   & 0.082                          \\
	2.210     & 62.79           & 2.013                   & 0.065                          \\
	2.210     & 65.39           & 1.819                   & 0.088                          \\
	2.210     & 67.99           & 1.722                   & 0.087                          \\
	2.210     & 70.59           & 1.355                   & 0.062                          \\
	2.210     & 73.20           & 1.318                   & 0.085                          \\
	2.210     & 75.80           & 1.160                   & 0.055                          \\
	2.210     & 78.40           & 0.968                   & 0.052                          \\
	2.210     & 81.00           & 0.999                   & 0.069                          \\
	2.210     & 83.60           & 0.727                   & 0.057                          \\
	2.210     & 86.20           & 0.683                   & 0.028                          \\
	2.210     & 88.81           & 0.600                   & 0.055                          \\
	2.210     & 91.41           & 0.500                   & 0.040                          \\
	2.210     & 94.01           & 0.657                   & 0.043                          \\
	2.210     & 96.61           & 0.507                   & 0.034                          \\
	2.210     & 99.21           & 0.264                   & 0.045                          \\
	2.210     & 101.82          & 0.263                   & 0.031                          \\
	2.210     & 104.42          & 0.273                   & 0.043                          \\
	2.210     & 107.02          & 0.295                   & 0.037                          \\
	2.210     & 109.62          & 0.279                   & 0.042                          \\
	2.210     & 112.22          & 0.055                   & 0.034                          \\
	2.210     & 117.43          & 0.141                   & 0.035                          \\
	2.210     & 120.03          & 0.153                   & 0.049                          \\
	2.773     & 1.47            & 14.086                  & 0.244                          \\
	2.773     & 4.07            & 18.717                  & 0.451                          \\
	2.773     & 6.67            & 20.572                  & 0.457                          \\
	2.773     & 9.27            & 23.100                  & 0.369                          \\
	2.773     & 11.88           & 29.789                  & 0.926                          \\
	2.773     & 13.18           & 29.649                  & 0.790                          \\
	2.773     & 15.78           & 32.012                  & 0.391                          \\
	2.773     & 18.38           & 33.119                  & 0.472                          \\
	2.773     & 20.98           & 34.282                  & 0.394                          \\
	2.773     & 23.58           & 29.139                  & 0.404                          \\
	2.773     & 27.49           & 34.813                  & 0.516                          \\
	2.773     & 30.09           & 30.211                  & 0.398                          \\
	2.773     & 32.69           & 28.690                  & 0.446                          \\
	2.773     & 35.29           & 26.919                  & 0.318                          \\
	2.773     & 37.89           & 17.906                  & 1.515                          \\
	2.773     & 39.19           & 18.713                  & 0.673                          \\
	2.773     & 41.80           & 16.163                  & 0.470                          \\
	2.773     & 44.40           & 13.844                  & 0.472                          \\
	2.773     & 47.00           & 11.048                  & 0.406                          \\
	2.773     & 49.60           & 7.855                   & 0.777                          \\
	2.773     & 53.50           & 8.313                   & 0.180                          \\
	2.773     & 56.11           & 7.290                   & 0.158                          \\
	2.773     & 58.71           & 5.847                   & 0.107                          \\
	2.773     & 61.31           & 4.909                   & 0.122                          \\
	2.773     & 63.91           & 4.836                   & 0.155                          \\
	2.773     & 66.51           & 3.345                   & 0.099                          \\
	2.773     & 67.81           & 2.693                   & 0.118                          \\
	2.773     & 70.42           & 2.199                   & 0.096                          \\
	2.773     & 73.02           & 1.785                   & 0.074                          \\
	2.773     & 75.62           & 1.491                   & 0.125                          \\
	2.773     & 79.52           & 1.650                   & 0.048                          \\
	2.773     & 82.12           & 1.655                   & 0.054                          \\
	2.773     & 84.73           & 1.201                   & 0.043                          \\
	2.773     & 87.33           & 1.380                   & 0.050                          \\
	2.773     & 89.93           & 0.884                   & 0.059                          \\
	2.773     & 92.53           & 1.075                   & 0.058                          \\
	2.773     & 93.83           & 0.869                   & 0.048                          \\
	2.773     & 96.43           & 0.898                   & 0.040                          \\
	2.773     & 99.04           & 0.719                   & 0.045                          \\
	2.773     & 101.64          & 0.729                   & 0.050                          \\
	2.773     & 105.54          & 0.675                   & 0.040                          \\
	2.773     & 108.14          & 0.579                   & 0.030                          \\
	2.773     & 110.74          & 0.719                   & 0.052                          \\
	2.773     & 113.35          & 0.461                   & 0.038                          \\
	2.773     & 115.95          & 0.482                   & 0.035                          \\
	2.773     & 118.55          & 0.478                   & 0.023                          \\
	2.773     & 119.85          & 0.370                   & 0.035                          \\
	2.773     & 122.45          & 0.369                   & 0.033                          \\
	3.128     & 1.95            & 16.149                  & 0.552                          \\
	3.128     & 4.55            & 25.038                  & 1.342                          \\
	3.128     & 7.15            & 28.630                  & 1.206                          \\
	3.128     & 9.75            & 31.324                  & 1.406                          \\
	3.128     & 12.35           & 34.984                  & 1.202                          \\
	3.128     & 14.96           & 31.955                  & 1.045                          \\
	3.128     & 17.56           & 36.798                  & 1.426                          \\
	3.128     & 20.16           & 37.381                  & 2.089                          \\
	3.128     & 22.76           & 35.313                  & 1.388                          \\
	3.128     & 25.36           & 31.631                  & 1.201                          \\
	3.128     & 27.96           & 33.364                  & 1.724                          \\
	3.128     & 30.57           & 24.693                  & 0.650                          \\
	3.128     & 33.17           & 20.735                  & 0.662                          \\
	3.128     & 35.77           & 18.137                  & 0.408                          \\
	3.128     & 38.37           & 14.648                  & 0.316                          \\
	3.128     & 40.97           & 16.378                  & 1.207                          \\
	3.128     & 43.58           & 10.473                  & 0.291                          \\
	3.128     & 46.18           & 8.265                   & 0.190                          \\
	3.128     & 48.78           & 7.276                   & 0.155                          \\
	3.128     & 51.38           & 5.410                   & 0.136                          \\
	3.128     & 53.98           & 5.456                   & 0.413                          \\
	3.128     & 56.58           & 3.422                   & 0.080                          \\
	3.128     & 59.19           & 2.859                   & 0.119                          \\
	3.128     & 61.79           & 2.636                   & 0.101                          \\
	3.128     & 64.39           & 2.203                   & 0.100                          \\
	3.128     & 66.99           & 1.976                   & 0.162                          \\
	3.128     & 69.59           & 1.764                   & 0.065                          \\
	3.128     & 72.20           & 1.638                   & 0.073                          \\
	3.128     & 74.80           & 1.375                   & 0.063                          \\
	3.128     & 77.40           & 1.317                   & 0.088                          \\
	3.128     & 80.00           & 1.323                   & 0.096                          \\
	3.128     & 82.60           & 1.115                   & 0.091                          \\
	3.128     & 85.20           & 1.134                   & 0.081                          \\
	3.128     & 87.81           & 0.809                   & 0.065                          \\
	3.128     & 90.41           & 0.801                   & 0.072                          \\
	3.128     & 93.01           & 0.760                   & 0.078                          \\
	3.128     & 95.61           & 1.011                   & 0.087                          \\
	3.128     & 98.21           & 0.954                   & 0.097                          \\
	3.128     & 100.82          & 0.912                   & 0.103                          \\
	3.128     & 103.42          & 0.806                   & 0.126                          \\
	3.128     & 106.02          & 0.387                   & 0.043                          \\
	3.128     & 108.62          & 0.384                   & 0.104                          \\
	3.128     & 111.22          & 0.412                   & 0.117                          \\
	3.128     & 113.82          & 0.773                   & 0.072                          \\
	3.128     & 116.43          & 0.516                   & 0.125                          \\
	3.128     & 119.03          & 0.154                   & 0.063                          \\* \bottomrule
\end{longtable}

\end{document}